\documentclass[12pt]{article}

\usepackage{amsmath,amsfonts,amssymb}
\usepackage{graphicx}
\usepackage{psfrag}
\usepackage{enumerate}
\usepackage{epsfig}
\usepackage{color}

\makeatletter \@addtoreset{equation}{section}

\makeatletter\renewcommand\section{\@startsection {section}{1}{\z@}%
                                   {-3.5ex \@plus -1ex \@minus -.2ex}
                                   {2.3ex \@plus.2ex}%
                                   {\normalfont\large\bfseries}}
\renewcommand\subsection{\@startsection{subsection}{2}{\z@}%
                                     {-3.25ex\@plus -1ex \@minus -.2ex}%
                                     {1.5ex \@plus .2ex}%
                                     {\normalfont\bfseries}}

\newcommand{\be}{\begin{equation}}
\newcommand{\ee}{\end{equation}}
\newcommand{\beq}{\begin{eqnarray}}
\newcommand{\eeq}{\end{eqnarray}}

\def\[{\left [}
\def\]{\right ]}
\def\({\left (}
\def\){\right )}

\def\r2{\sqrt{2}}

\newcommand{\bbibitem}[1]{\bibitem{#1}\marginpar{#1}}

\def\Label#1{\label{#1}%
  \smash{\hbox to0pt{\raise1ex\hbox{\tiny[#1]}\hss}}}
\def\noLabels{\let\Label=\label}
\def\nobbibitem{\let\bbibitem=\bibitem}

\long\def\symbolfootnote[#1]#2{\begingroup%
\def\thefootnote{\fnsymbol{footnote}}\footnote[#1]{#2}\endgroup}

\begin{document}
\noLabels 
\nobbibitem 

\begin{titlepage}


\vfil\

\begin{center}

{\Large \bf The grainy multiverse}

\vspace{12mm}

{\large Bart{\l}omiej Czech\symbolfootnote[1]{\tt email:czech@phas.ubc.ca}}
\\

\vspace{5mm}

\bigskip\medskip
\bigskip\centerline{\it Department of Physics and Astronomy}
\smallskip\centerline{\it University of British Columbia}
\smallskip\centerline{\it 6224 Agricultural Road}
\smallskip\centerline{\it Vancouver, BC V6T 1Z1}
\smallskip\centerline{\it Canada}


\vfil

\end{center}
\setcounter{footnote}{0}
\begin{abstract}
\noindent
I consider a landscape containing three vacua and study the topology of global spacelike slices in eternal inflation. A discrete toy model, which generalizes the well studied Mandelbrot model, reveals a rich phase structure. Novel phases include monochromatic tubular phases, which contain crossing curves of only one vacuum, and a democratic tubular phase, which contains crossing curves of all three types of vacua. I discuss the generalization to realistic landscapes consisting of many vacua. Generically, the system ends up in a grainy phase, which contains no crossing curves or surfaces and consists of packed regions of different vacua. Other topological phases arise on the scale of several generations of nucleations. 

\end{abstract}
\vspace{0.5in}

\end{titlepage}
\renewcommand{\baselinestretch}{1.05}  

\newpage

\section{Introduction}

The realization that string theory gives rise to a landscape of vacua opened the possibility that our universe may be one of many nucleating bubbles expanding in an eternally inflating false vacuum \cite{bp, landscape}. The resulting picture poses a number of novel, conceptual problems: it is difficult to consistently define probabilities in eternal inflation (the cosmological measure problem) while the richness of the string landscape may pose an intrinsically insurmountable obstacle to phenomenological studies traceable to complexity theory \cite{complexity}. This paper follows a suggestion of \cite{winitzki} to study one aspect of eternal inflation that is not afflicted by these difficulties, namely the topology formed by the nucleating bubbles. One such study was carried out in \cite{susskind}, whose authors considered the topology and physics arising from a toy landscape containing two vacua, called white and black. This paper explores the effects of including more realistic, richer landscapes on the resulting multiverse topologies.

Classifying multiverses by topology is worthwhile for several reasons. First, it is well defined: 
I will be looking at topologies formed by nucleating and expanding bubbles on global spacelike slices of an inflating geometry, but as shown in \cite{susskind}, these topologies are independent of the choice of slicing. Second, topology allows one to study the global picture of eternal inflation without tackling the cosmological measure problem first. Third, it is robust in that it is not sensitive to details of bubble wall dynamics, collisions or other modeling peculiarities. Consequently, one can reliably resort to simplified models; in what follows I shall employ a discrete toy model due to Mandelbrot \cite{mandelbrot}, which has beeen utilized previously in studies of eternal inflation, notably in \cite{susskind}. Finally, the different topological phases of eternal inflation are characterized by different physics, much of which has not yet been fully understood. In Sec.~\ref{prelim} I will briefly review the results of \cite{susskind}, who gave a physical characterization of the topological phases arising from a two-vacuum landscape. It is important to understand which of their findings carry over to more realistic, richer landscapes. As a first step in this direction, the present paper classifies the topologies arising from a landscape consisting of three vacua (called white, grey and black) and discusses the many-vacuum limit qualitatively.

Topological phases of eternal inflation are differentiated by the existence of crossing curves and crossing surfaces \cite{susskind}. Roughly, the system is said to possess white crossing curves if at arbitrarily late times, any two arbitrarily far regions are connected by a curve wholly surrounded by the white vacuum. The definition of white crossing surfaces pertains to the existence at arbitrarily late times of surfaces entirely contained in the white region, whose boundary is arbitrarily large. Of course, the definitions for grey and black crossing curves and surfaces are analogous. Below I consistently use abbreviations WXC (White X-ing Curves) and WXS (White X-ing Surfaces) and their analogues for Grey and Black regions. As shown in \cite{susskind}, the existence of crossing curves and surfaces is independent of the choice of time slicing. Topological phases are defined over a parameter space, whose natural coordinates are the dimensionless nucleation rates of the bubbles of various colors, $\gamma = \Gamma H^{-4}$, where $H^{-4}$ is the Hubble volume.

The main results of this paper are phase diagrams, which describe how multiverse topology varies over the space of landscape parameters. I consider explicitly a landscape containing three vacua and comment on the general case in the Discussion. In finding the phase diagrams, I employ the following strategy:
\begin{enumerate} 
\item Reduce the problem to its discrete version, a three-vacuum analogue of the Mandelbrot model \cite{mandelbrot, susskind}. This, along with a brief review of previous results, is the content of Sec.~\ref{prelim}.
\item Identify sectors in the parameter space of the discrete model, whose phases can be inferred from the phase structure of the well studied two-vacuum Mandelbrot model. This is the content of Sec.~\ref{ingredients}.
\item Assemble the full phase diagrams (Secs.~\ref{2ddiag}-\ref{3ddiag}).
\end{enumerate}
This methodology qualitatively determines the phase structure of the three-vacuum Mandelbrot model. Though one cannot lift it to continuous eternal inflation on a quantitative level, its qualitative lessons are robust and generalizable to more realistic landscapes. They suggest that rich landscapes generically give rise to a grainy phase, in which the geometry is packed with a hodgepodge of bubbles of various colors. Other topological phases arise on the scale of several generations of nucleations. More details are given in the Discussion.

\section{Preliminaries}
\Label{prelim}

\subsection{Setup}
\Label{setup}

Consider a toy landscape consisting of three vacua: a false vacuum labeled white, an intermediate vacuum labeled grey, and a true vacuum labeled black. Assume that each vacuum has a non-negative cosmological constant. Start in an initial all-white universe. 
Bubbles of grey or black vacuum will form in the white region \cite{cdl} with rates $\Gamma^{\rm wg}$ and $\Gamma^{\rm wb}$ and further, inside the nucleated grey regions, bubbles of black vacuum will appear with the rate $\Gamma^{\rm gb}$. I assume that all transitions are irreversible (e.g. $\Gamma^{\rm wg}>0 \Rightarrow \Gamma^{\rm gw}=0$), but otherwise the nucleation rates are not subject to any constraints. The $\Gamma$'s compete with the rates of expansion of the parent white and grey vacua: when $\Gamma H^{-4} \gtrsim 1$, the nucleated regions take over the geometry, but when this condition is not satisfied, the white (or grey) region persists indefinitely, giving rise to eternal inflation. The present paper seeks to understand the connectivity of grey and black regions on spatial slices of this system at late times, as functions of the dimensionless nucleation rates $\gamma^{\rm wb} \equiv \Gamma^{\rm wb} H_{\rm w}^{-4}$ and their analogues $\gamma^{\rm wg}$ and $\gamma^{\rm gb}$. Ref.~\cite{susskind} showed that this question is well defined, because the topology of the regions filled with descendant vacua is independent of the choice of time slicing.

\subsection{Topological phases of eternal inflation with two vacua}
\label{2dreview}

Ref.~\cite{susskind} (see also \cite{futurefoam}) studied the analogous problem for a landscape consisting of two vacua. They found four topological phases, distinguished by the presence or absence of white crossing surfaces (WXS), white crossing curves (WXC) and their black analogues (BXS and BXC). Formally, these objects are defined as follows. Consider a three-dimensional ball of constant comoving radius in the multiverse, and select two disjoint open sets on the sphere surrounding it. If at arbitrarily late times the ball possesses a curve entirely contained in regions of one vacuum (definite color, henceforth `monochromatic') that connects the two open sets, then we say that the multiverse contains crossing curves of that color. Likewise, if at arbitrarily late times the ball contains a monochromatic surface that bisects it and screens the two open sets from one another, then we say that the multiverse contains crossing surfaces of that color. 

The four topological phases identified by \cite{susskind} are:
\begin{enumerate} 
\item {\bf The Black Island Phase}, which contains WXS but no BXC or BXS. The white regions inflate eternally. The black regions are isolated CdL bubbles filled with open FRW universes. They will occassionally collide with other similar bubbles so that their boundary will be a surface whose genus increases in time.
\item {\bf The Tubular Phase}, which contains crossing curves of both colors, but no crossing surfaces of either color. It is separated from the black island phase by a percolation transition studied in \cite{guthweinberg}. The genus of the boundary of the black region becomes infinite in finite time. 
\item {\bf The White Island Phase} contains BXS but no WXC or WXS. It is characterized by a phenomenon, which \cite{susskind} called `cracking,' whereby the white regions become disconnected. The analysis of \cite{susskind} indicates that this leads to a formation of singularities in black regions, which may be mimicked by Kruskal-Schwarzschild black holes. Thus, an observer in a black region does not see all the disconnected boundaries, but rather a finite number of boundaries demarcating white regions and a finite number of black hole horizons, each sized no larger than roughly the Hubble scale of the ancestor vacuum. A likely final fate of this geometry is a big crunch.
\item {\bf The Aborted Phase} is the phase in which the black vacuum entirely takes over the geometry while the white regions disappear. This takes place when $\Gamma^{\rm wb} H_{\rm w}^{-4} \gtrsim 1$. 
\end{enumerate}
Part of the motivation of this work is to check to what degree these findings extend to more general scenarios. As a first step toward that goal, I classify the phases of the three-vacuum Mandelbrot model by their topology, leaving their observational or theoretical characterization to future work. 

\subsection{Deriving the Mandelbrot model}
\label{derivation}

Ref.~\cite{susskind} motivated and illustrated their findings with a discrete model of eternal inflation. Variants of the same model were also used in prior studies of eternal inflation \cite{guthweinberg, aryalvilenkin, winitzki, timerep}. Because the present paper deals with a more complex situation, it is useful to carefully derive the correspondence between eternal inflation and its discrete analogue.

The interesting topology of eternal inflation arises from collisions between bubbles. A bubble of a descendant vacuum grows with a speed approaching the speed of light \cite{cdl}. Meanwhile, the surrounding parent vacuum expands with a rate determined by its Hubble scale. Thus, in comoving coordinates of the parent vacuum, the bubble approaches a finite size given by the Hubble scale of the parent at the time of nucleation. Consequently, in cosmologies characterized by an accelerated expansion such as de Sitter, younger bubbles are smaller {\it when expressed in coming coordinates}. This is represented on the left of Fig.~\ref{discretizing}.

\begin{figure}[t]
\begin{center}
\vspace{-0.3in}
\includegraphics[scale=0.34]{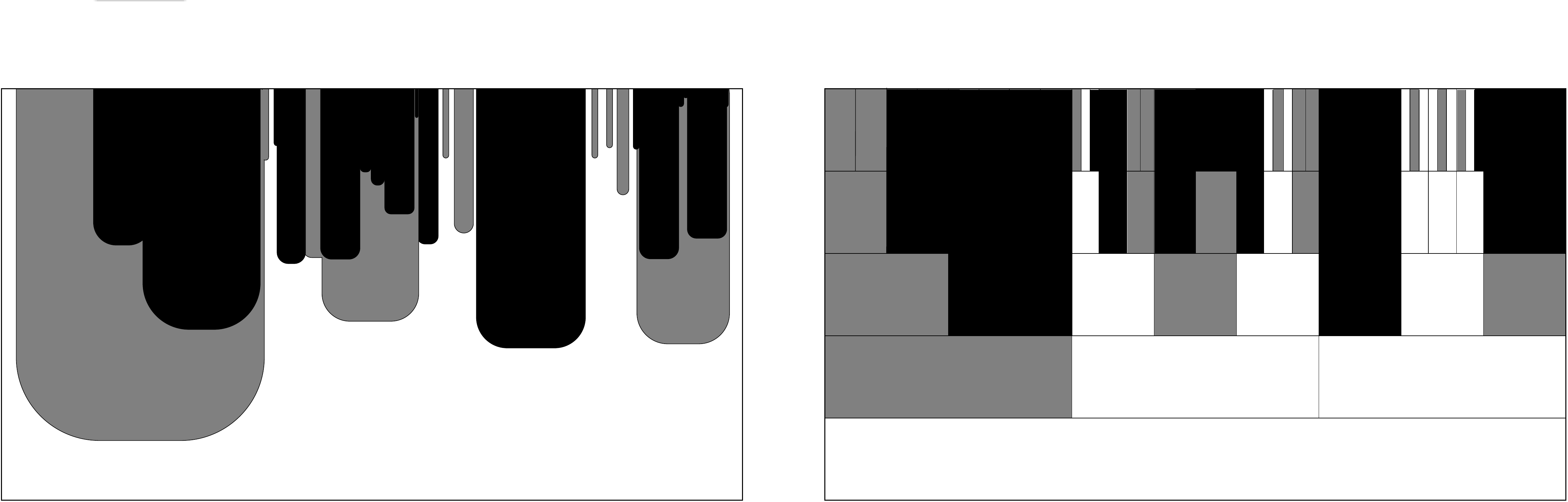}
\caption{An example history of three-vacuum eternal inflation (left) and its discretized analogue (right). Horizontal axes correspond to comoving coordinates and span several initial Hubble lengths; the vertical axes mark time. The discrete evolution was emulated for $p_{\rm wb} = 1/10,\, p_{\rm wg} = 1/3,\, p_{\rm gb} = 1/2,\, N_{\rm w}=3,\, N_{\rm g} = 2$.}
\Label{discretizing}
\end{center}
\end{figure}

The goal is to study this system using a discrete model. How to discretize the left of Fig.~\ref{discretizing}? The first step is to chop the spatial slices of the geometry into discrete cells. The optimal cell size is the Hubble scale of the ancestor vacuum. With this choice, a nucleation event can be represented by filling the cell with the color of the nucleated vacuum. This is consistent, because each bubble quickly attains the parent Hubble size, though it never outgrows it. Furthermore, two bubbles are expected to collide if their nucleations are separated by distances smaller than the parent Hubble distance; otherwise their growth does not catch up with the expansion of the surrounding vacuum and they remain disconnected. This rule, which determines the connectivity of the bubbles, is naturally captured by setting the spatial unit cell to be Hubble sized.

The Hubble length in comoving coordinates is given by $1/\dot{a}(t)$, where $a(t)$ is the scale factor. If the expansion of the parent vacuum accelerates, $1/\dot{a}(t)$ is a decreasing function of time, which means that the Hubble length, when expressed in comoving coordinates, decreases. Thus, our unit spatial cells must in time be subdvided into smaller units. If one splits time into $\Delta t$ intervals, each unit cell must be subdivided into
\begin{equation}
\label{relationn}
N = \frac{\dot{a}(t + \Delta t)}{\dot{a}(t)}
\end{equation}
subunits in each linear dimension. Thus, in three spatial dimensions, one subdivides each unit cell into $N^3$ sub-cells while in two dimensions the subdivision is into $N^2$ subunits. To take advantage of the convenience of a discrete model, one should select $\Delta t$ so that $N$ is a natural number. In de Sitter, this corresponds to setting $\Delta t$ to be an order unity multiple of the Hubble time. The resulting model is known as the Mandelbrot model \cite{mandelbrot}.

It remains to find a discrete analogue for the nucleation rate $\Gamma$, which defines the probability with which bubbles nucleate in a given four-volume. Our time intervals and spatial cells select a discrete unit of four-volume $H^{-3} \Delta t$ and, by extension, a dimensionless quantity $\Gamma H^{-3} \Delta t$. The probability with which the cells of the discrete model are colored black (or grey) must be a function of this quantity:
\begin{equation} 
\label{relationp}
p = p\, (\Gamma H^{-3} \Delta t)
\end{equation}
Setting $p = \Gamma H^{-3} \Delta t$ is natural, but it is not the only reasonable choice. For example, the frequency of bubble nucleations in a given four-volume is given by the Poisson distribution; from that point of view, {\it not coloring} a discrete cell should happen with probability $1-p \to \exp{(-\Gamma H^{-3} \Delta t)}$. The ambiguity in defining $p$ shows that one should not attempt to derive quantitative results about eternal inflation from the Mandelbrot model. It is only a toy model of eternal inflation, which is expected to carry qualitative lessons.

I close this subsection by highlighting the definition of the Mandelbrot model in the form, in which I will use it below. Begin with an infinite lattice of white cubes\footnote{I consider an infinite lattice instead of a single cube to eliminate tedious artifacts of an infrared cutoff.} and iterate the following procedure: subdivide each white cube into $N_{\rm w}^3$ ($N_{\rm w}^2$ in two dimensions) smaller cubes and color them grey with probability $p_{\rm wg}$ and black with probability $p_{\rm wb}$; likewise, subdivide each grey cube into $N_{\rm g}^3$ ($N_{\rm g}^2$ in two dimensions) smaller cubes and color them black with probability $p_{\rm gb}$. Take the limit of infinitely many iterations. Check for the existence of monochromatic curves that span an infinite distance (crossing curves) and for the existence of monochromatic two-surfaces that are unbounded in any direction (crossing surfaces). Note that $N_{\rm w} \neq N_{\rm g}$, because the expansions of the white and grey vacua accelerate at different rates (see eq.~\ref{relationn}). 

\subsection{Topological phases of the Mandelbrot model with two vacua}
\label{2dmandelbrot}

The phase structure of the white-black Mandelbrot model in three dimensions was established in \cite{chayes1, chayes2}. It mimics the phases of eternal inflation reviewed in Sec.~\ref{2dreview}. There is a black island phase (which contains WXS and no BXC), a tubular phase (which has crossing curves of both colors but no crossing surfaces), a white island phase (which contains BXS and no WXC), and the aborted phase (all black). The crossover value between the black island phase and the tubular phase is denoted $p_{\rm s}$: after infinitely many iterations, BXC appear and WXS are eliminated if and only if $p \geq p_{\rm s}$. In the tubular phase white and black regions each form connected networks of infinite genus. The critical value separating the tubular and the white island phase is $p_{\rm c}$: after infinitely many iterations BXS appear and WXC are eliminated if and only if $p \geq p_{\rm c}$. The phase transitions at $p_{\rm s},\,p_{\rm c}$ are first order.\footnote{At least in some models of eternal inflation, one would anticipate transitions that are second or higher order \cite{freivogel2}. This disagreement is not relevant to the purposes of this paper, which are qualitative.} Finally, in the aborted phase every region becomes all black after sufficiently many iterations. The onset of the aborted phase happens at $p_\emptyset(N) = 1 - N^{-3}$. The other transition points also depend on $N$. The exact forms of $p_{\rm s}(N),\,p_{\rm c}(N)$ as functions of $N$ are not known.

I will also study the two-dimensional Mandelbrot model, that is one defined over a lattice of squares instead of cubes and in which cells are divided into $N^2$ units at each iteration. It is a toy model for studying the topology of grey and black regions on a spacelike two-dimensional surface, such as the wall surrounding a fiducial CdL bubble at a late time (compare with \cite{freivogel2}). The two-dimensional model contains three phases: the black island phase (with WXC but no BXC), the white island phase (with BXC but no WXC) and the aborted phase. I will denote the boundary between the black island and the white island phases with $p_{\rm c}(N)$. In two dimensions $p_\emptyset(N) = 1 - N^{-2}$.

\section{Phase diagrams}
\Label{generalization}

The goal is to understand the phase structure of three-vacuum eternal inflation (Sec.~\ref{setup}). I follow the strategy outlined in the Introduction. The first step -- a discretization of the system -- was accomplished in Sec.~\ref{derivation}. In this section the next two steps of the strategy are carried out. Sec.~\ref{ingredients} identifies special loci in the parameter space of the three-vacuum Mandelbrot model, whose topological phases can be inferred from the two-vacuum model and its phase structure (reviewed in Sec.~\ref{2dmandelbrot}). These ingredients are then assembled to form the full phase diagrams of the three-vacuum Mandebrot model in two (Sec.~\ref{2ddiag}) and three dimensions (Sec.~\ref{3ddiag}). I will discuss the extent to which these results can be lifted back to realistic, continuous eternal inflation in the Discussion. 

Before proceeding, some comments are in order. In the two-vacuum Mandelbrot model, the phases are characterized by whether at late enough times the white region contains crossing surfaces, crossing curves, or whether it is empty. The threshold values $p_{\rm s}, p_{\rm c}, p_\emptyset$ are functions of $N$, the number of subdivisions of white cells at each iteration. The technique is to lift these results to the three-vacuum model whenever possible. However, the three-vacuum Mandelbrot model contains two different values of $N$: $N_{\rm w}$, which defines the number of subdivisions of white cells, and $N_{\rm g}$, which is the number of subdivisions of grey cells. Consequently, the threshold values for the existence of white and grey crossing curves and surfaces, as well as for the white and grey regions to be empty, are different from one another. 
For this reason, I use superscripts to distinguish the critical probabilities governing the behavior of white and grey regions. For example, the probability of finding a WXS after infinitely many iterations vanishes if and only if $p_{\rm wg} + p_{\rm wb} \geq p_{\rm s}^{\rm w}$, but an analogous statement for grey regions, whatever its exact form, must depend on $p_{\rm s}^{\rm g}$ and not $p_{\rm s}^{\rm w}$. Note that the parameter space of the three-vacuum Mandelbrot model is a product of a 1-simplex ($0 \leq p_{\rm gb} \leq 1$) and a 2-simplex ($0 \leq p_{\rm wg} + p_{\rm wb} \leq 1$). I do not consider variations of $N_{\rm w}$ and $N_{\rm g}$ explicitly, because their effect is captured by shifting the phase boundaries $p_{\rm s}^{\rm w}, p_{\rm s}^{\rm g}, p_{\rm c}^{\rm w}, p_{\rm c}^{\rm g}, p_\emptyset^{\rm w}, p_\emptyset^{\rm g}$.

In all figures in this section, phase divisions that are known exactly (as functions of $p_{\rm s}^{\rm w}(N_{\rm w})$, etc.) are distinguished with the use of continuous lines from those which are known only qualitatively, represented by dashed lines. Dotted lines are auxiliary.

\subsection{Ingredients}
\label{ingredients}
This subsection compiles the reasoning, by which results from the two-vacuum Mandelbrot model may be adapted to the three-vacuum case. Whenever possible, I list facts in indexed lists and use the same indices in the figures to mark the corresponding features of phase diagrams.
I consider in turn the reasoning applicable to white, black, and grey regions.

\paragraph{White Regions}
\begin{figure}[t]
\begin{center}
\includegraphics[scale=0.6]{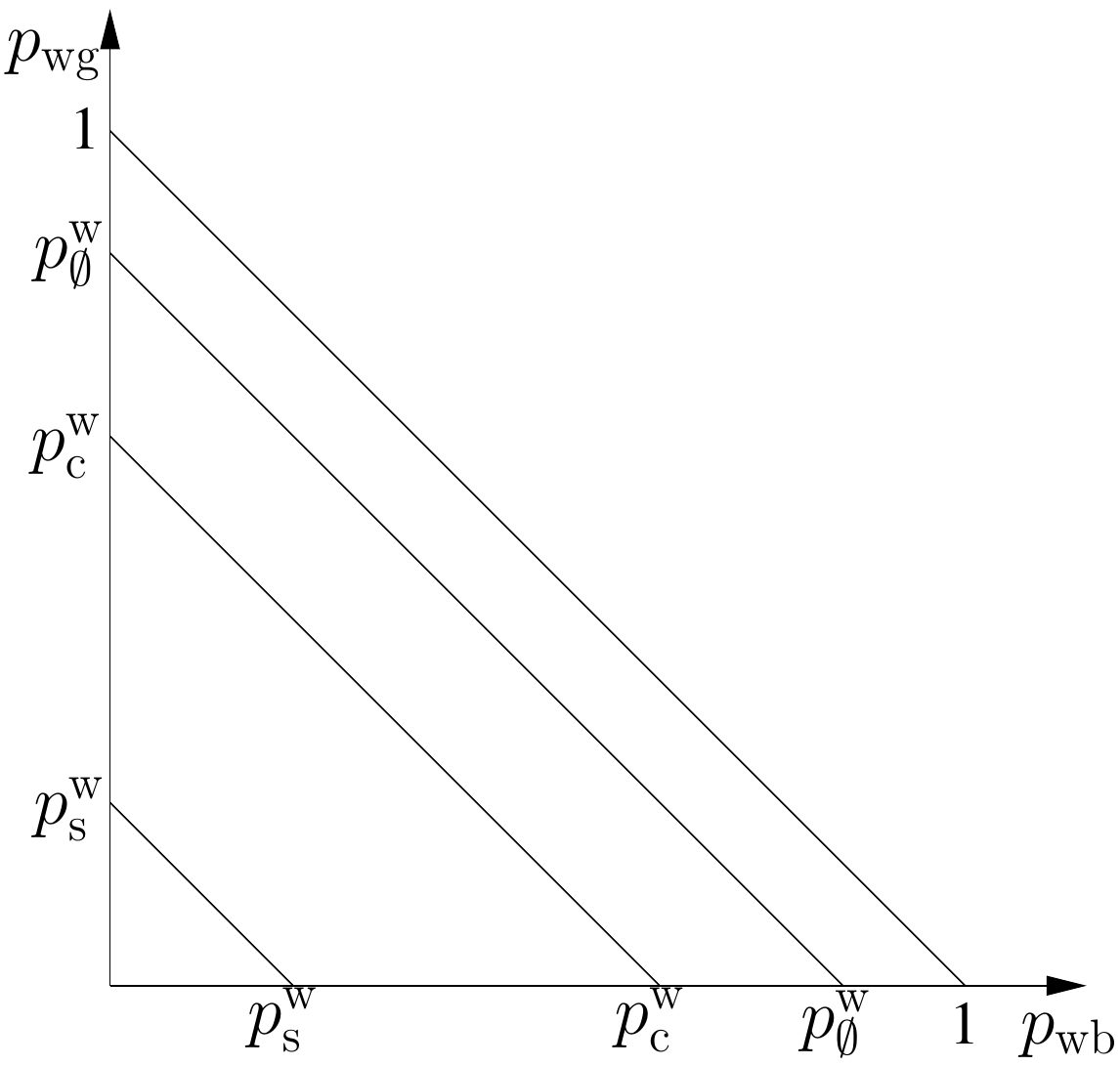}
\caption{Moving away from the origin, the successive regions contain WXS, WXC, white islands and no white at all (the white-aborted phase). The points marked on the $p_{\rm wb}$-axis, $p^{\rm w}_{\rm s},\,p^{\rm w}_{\rm c},\,p^{\rm w}_\emptyset$, are the locations of phase transitions in the two-vacuum Mandelbrot model.}
\Label{whitereg}
\end{center}
\end{figure}

For the purposes of studying the topology of the white regions, it is sufficient to focus on the projection of the parameter space onto the 2-simplex $0 \leq p_{\rm wg} + p_{\rm wb} \leq 1$, because $p_{\rm gb}$ plays no role here. On the projection, the different regimes are separated by downward-sloping 45$^\circ$ lines intersecting the $p_{\rm wb}$ axis at the critical values. The logic is that from the viewpoint of the white region, nucleations of the grey and black types are indistinguishable, so the boundary of the region containing WXS must be a level set of $p_{\rm wb} + p_{\rm wg}$. For example, since $p^{\rm w}_{\rm s}$ marks the two-vacuum cross-over from the black island (including WXS) to the tubular (including WXC) phase, the line $p_{\rm wg} + p_{\rm wb} = p^{\rm w}_{\rm s}$ demarcates the presence and absence of WXS. Analogous reasoning reveals the significance of parallel lines intersecting the values $p^{\rm w}_{\rm c}$ and $p^{\rm w}_\emptyset$, which separate the region containing only islands of white from that containing WXC and from the white-aborted phase. Facts about the white regions, which follow from the two-vacuum results, are summarized in Fig.~\ref{whitereg}.

\paragraph{Black Regions}
\begin{figure}[t]
\begin{center}
\includegraphics[scale=0.6]{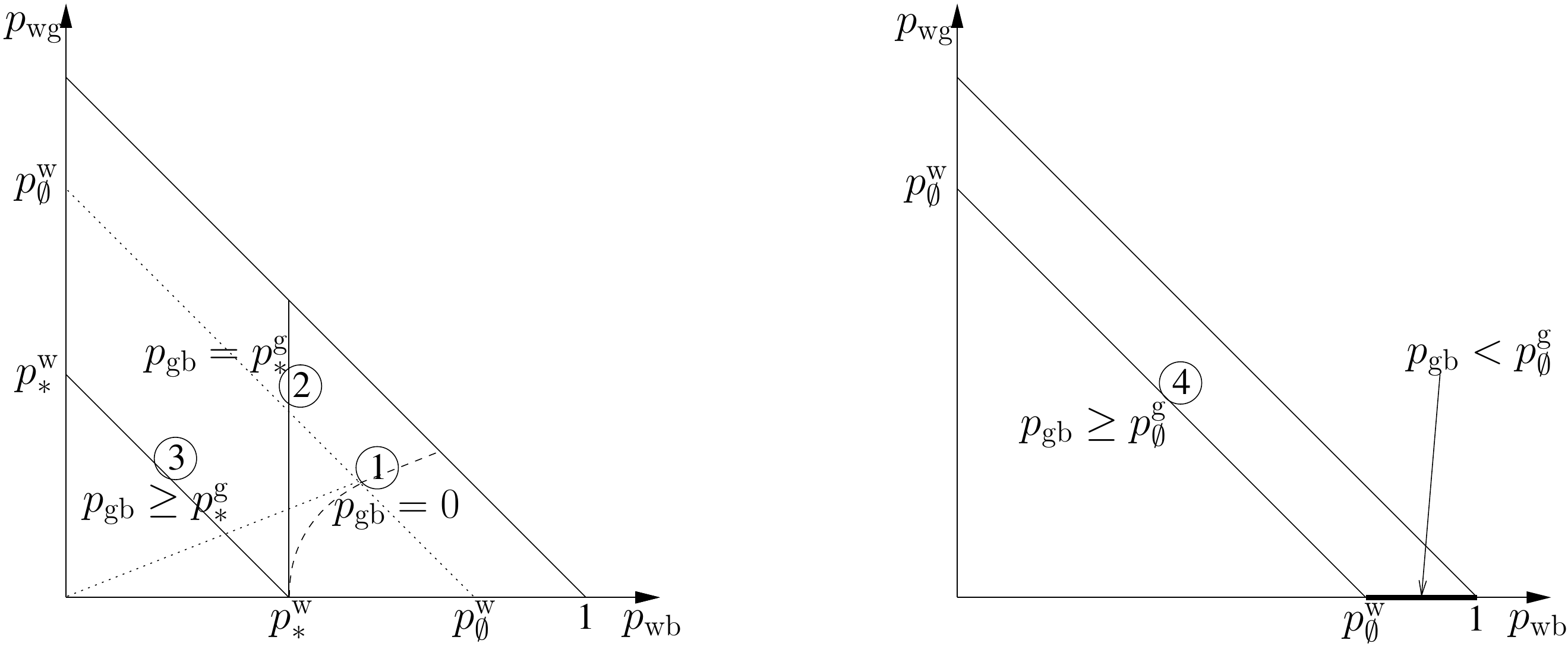}
\caption{The lines demarcating topological phases of black bubbles at constant $p_{\rm gb}$. Left: the boundaries of the regions containing BXC ($p^{\rm w}_{\rm *} = p^{\rm w}_{\rm s}$) or BXS ($p^{\rm w}_{\rm *} = p^{\rm w}_{\rm c}$). Right: the aborted (all black) phase. The numbers refer to the observations listed in the text.}
\Label{blackreg}
\end{center}
\end{figure}

The topology of the black regions cannot be simply read off from the two-vacuum results, because it depends heavily on the value of $p_{\rm gb}$. One may determine the boundaries between the topological phases of the black vacuum in the following steps, indexed in Fig.~\ref{blackreg}:
\begin{enumerate}
\item Consider BXS and BXC. For positive $p_{\rm wg}$ and at $p_{\rm gb} = 0$, their existence necessitates greater values of $p_{\rm wb}$ than in the two-vacuum system, because any white$\rightarrow$grey transitions effectively lock those regions from ever becoming black. Therefore, a phase boundary at constant $p_{\rm gb} = 0$ initially curves towards the diagonal $p_{\rm wb} + p_{\rm wg} = 1$, but becomes a straight line in the region $p_{\rm wb} + p_{\rm wg} \geq p^{\rm w}_\emptyset$, because in the white-aborted phase and in the absence of grey$\rightarrow$black transitions the relative abundance of grey and black can only depend on the ratio $p_{\rm wb} / p_{\rm wg}$.
\item Consider the boundary between the phases containing black islands and BXC as $p_{\rm gb}$ increases. When $p_{\rm gb} = p^{\rm g}_{\rm s}$, an intermediate formation of grey regions has no effect on the eventual formation of BXC, so that boundary is vertical (independent of $p_{\rm wg}$). When $p_{\rm gb} = p^{\rm g}_{\rm c}$, an analogous argument holds for the boundary between the phases containing BXC and BXS.
\item If $p_{\rm gb} \geq p^{\rm g}_\emptyset$, then every grey region eventually becomes black and $p_{\rm wb} + p_{\rm wg}$ is again the relevant quantity for studying the topology of black. Point~3 of Grey Regions extends this conclusion to $p_{\rm gb} \geq p^{\rm g}_{\rm s},\, p^{\rm g}_{\rm c}$ (for BXC and BXS, respectively). 
\item Finally for the aborted phase, in addition to the condition inherited from the two-vacuum system, $p_{\rm wb} \geq p^{\rm w}_\emptyset$, for positive $p_{\rm wg}$ one must also impose $p_{\rm gb} \geq p^{\rm g}_\emptyset$ to ensure that all the nucleating grey regions eventually become black.
\end{enumerate}

\paragraph{Grey Regions}
\begin{figure}[t]
\begin{center}
\includegraphics[scale=0.5]{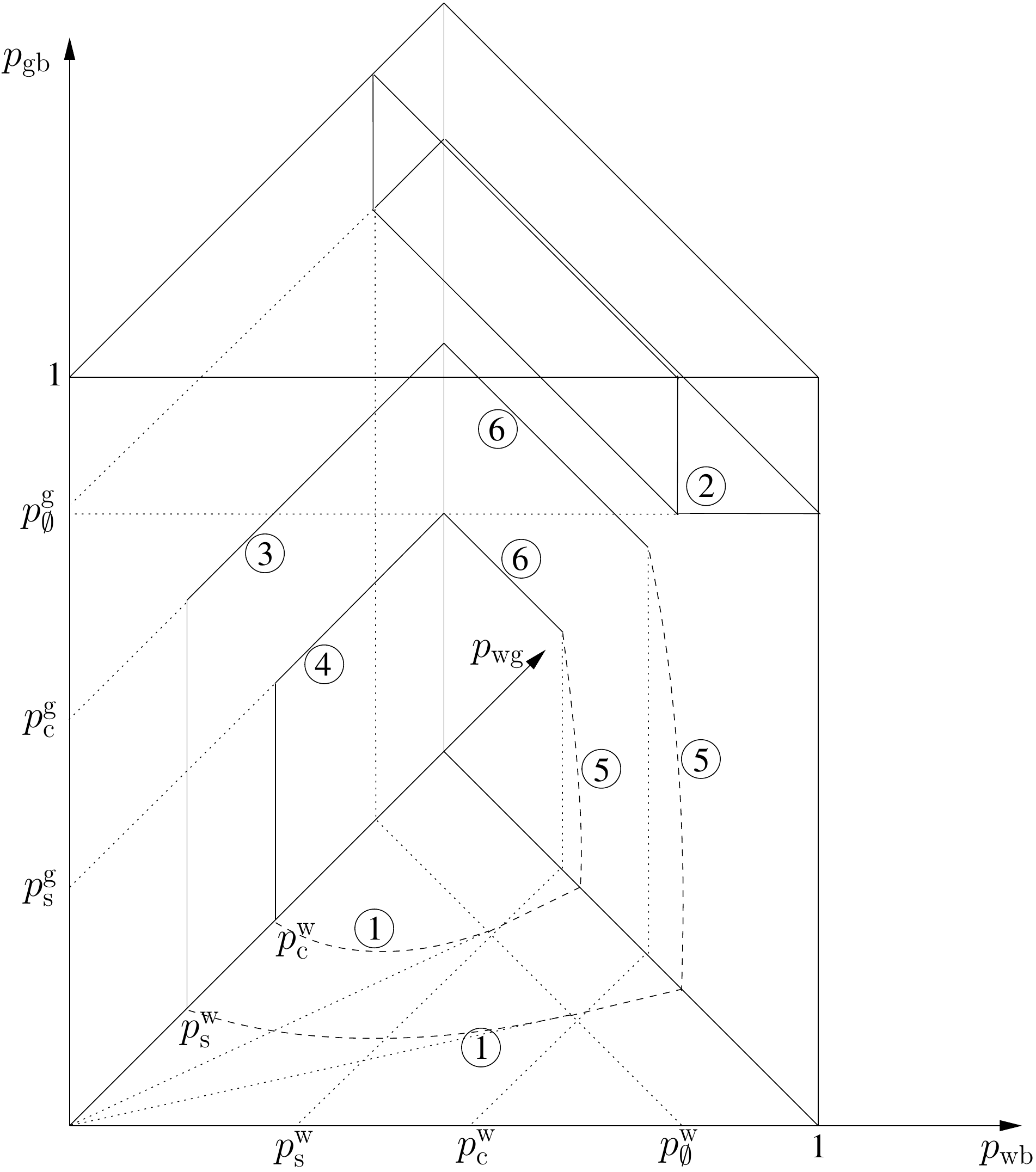}
\caption{The topological phases of the grey regions. Moving along the $p_{\rm wg}$-axis away from the origin, the successive domains contain grey islands, GXC, and GXS. The top wedge of the diagram is the aborted phase without any grey regions. The numbers refer to the observations listed in the text.}
\Label{greyreg}
\end{center}
\end{figure}
The topology of the grey regions is determined by the following facts, which are indexed in Fig.~\ref{greyreg}:
\begin{enumerate}
\item For $p_{\rm gb} = 0$, the phase diagram is grey$\leftrightarrow$black symmetric and Fig.~\ref{blackreg} describes GXS and GXC, too.
\item For positive $p_{\rm wg}$ a necessary and sufficient condition for the system to contain no grey regions is $p_{\rm wb} + p_{\rm wg} \geq p^{\rm w}_\emptyset$ and $p_{\rm gb} \geq p^{\rm g}_\emptyset$. To see this, note that the expected number of grey bubbles in one initial lattice cube after $i+1$ iteration steps is
    \begin{equation}
    \begin{array}{rcl}
    n_{\rm g}(i+1) & = & N_{\rm w}^d\, p_{\rm wg}\, \sum_{k=0}^{i} \Big(N_{\rm w}^d (1-p_{\rm wg} - p_{\rm wb}) \Big)^k \Big(N_{\rm g}^d (1-p_{\rm gb}) \Big)^{i-k} \\
    & = & N_{\rm w}^d\, p_{\rm wg}\, \frac{\big(N_{\rm w}^d (1-p_{\rm wg} - p_{\rm wb}) \big)^{i+1} - \big(N_{\rm g}^d (1-p_{\rm gb}) \big)^{i+1}}{\big(N_{\rm w}^d (1-p_{\rm wg} - p_{\rm wb}) \big) - \big(N_{\rm g}^d (1-p_{\rm gb}) \big)}\,,
    \end{array}\tag{4}
    \end{equation}
    where $k$ in the summation indexes the generation of grey bubbles that formed at the $(k+1)^{\rm th}$ step of the iteration. This expression cannot vanish unless both terms in the numerator go to zero individually, which is equivalent to the condition stated above. Physically, the requirement on $p_{\rm wb} + p_{\rm wg}$ is necessary because when it is violated, white regions continue to nucleate grey bubbles forever, which therefore persist arbitrarily late despite the fact that each individual grey bubble eventually turns entirely black.
\item Consider the regime $p_{\rm wb} = 0$ and $p_{\rm wg} \geq p^{\rm w}_{\rm s}$. If we ignore grey$\to$black nucleations, the system develops GXC. One may then emulate the effect of a non-zero $p_{\rm gb}$ by studying an auxiliary two-vacuum (white vs. non-white) Mandelbrot model, in which with probability $p_{\rm wg}$ one fills cells with grey cubes that already contain appropriate black structures inside them: black islands for $p_{\rm gb} < p^{\rm g}_{\rm s}$, black crossing curves (spanning the length of the cube) for $p^{\rm g}_{\rm s} \leq p_{\rm gb} < p^{\rm g}_{\rm c}$, or black crossing surfaces (spanning the cross section of the cube) for $p_{\rm gb} \geq p^{\rm g}_{\rm c}$. Only the last possibility represents an obstruction to GXC, which shows that on $p_{\rm wb} = 0$ and $p_{\rm wg} \geq p^{\rm w}_{\rm s}$, the boundary of the grey-tubular phase is $p_{\rm gb} = p^{\rm g}_{\rm c}$. When $p^{\rm g}_{\rm s} \leq p_{\rm gb} < p^{\rm g}_{\rm c}$, one obtains a grey tubular network of infinite genus forming GXC, which is percolated by a black tubular network of infinite genus containing BXC.  This confirms the statement anticipated in Point~3 of Black Regions -- that BXC exist so long as $p_{\rm wg} + p_{\rm wb} \geq p^{\rm w}_{\rm s}$ and $p_{\rm gb} \geq p^{\rm g}_{\rm s}$. An identical argument shows that BXS exist so long as $p_{\rm wg} + p_{\rm wb} \geq p^{\rm w}_{\rm c}$ and $p_{\rm gb} \geq p^{\rm g}_{\rm c}$.
\item One may similarly deduce the boundary of the phase containing GXS in the regime $p_{\rm wb} = 0$ and $p_{\rm wg} \geq p^{\rm w}_{\rm c}$. Again consider an auxiliary two-vacuum (white vs. non-white) Mandelbrot model, in which with probability $p_{\rm wg}$ one fills cells with grey cubes that already contain the black structures dictated by $p_{\rm gb}$. Now, as soon as $p_{\rm gb} \geq p^{\rm g}_{\rm s}$, the BXC spanning the grey cubes preclude the survival of GXS. Thus, on $p_{\rm wb} = 0$ and $p_{\rm wg} \geq p^{\rm w}_{\rm c}$, the boundary of the phase containing GXS is $p_{\rm gb} = p^{\rm g}_{\rm s}$.
\item Consider the boundaries of the loci containing grey islands only, GXC, and GXS, and fix attention on the region $p_{\rm wb} + p_{\rm wg} \geq p^{\rm w}_\emptyset$, where the system is all grey and black. By analogy with the two-vacuum system, these boundaries must coincide with phase boundaries demarcating domains with BXS, BXC, and black islands, respectively.
\item Next, we noted in Point~2 of Fig.~\ref{blackreg} that these boundaries contain the points 
\begin{equation*} \tag{5}
(p_{\rm wb},\,p_{\rm wg},\,p_{\rm gb}) = (p^{\rm w}_{\rm *},\,1-p^{\rm w}_{\rm *},\,p^{\rm g}_{\rm *}), 
\end{equation*}
with $p_{\rm *} = p_{\rm c},\,p_{\rm s}$, respectively. On the other hand, Points~3 and 4 above dictate that the same boundaries contain the points $(0,1,p_{\rm *}^{\rm g})$. These two pairs of points are connected with straight lines.
\end{enumerate}

\subsection{Three vacua in two dimensions}
\label{2ddiag}

It is interesting to study the topology of bubble collisions with the wall of a fiducial bubble. The Mandelbrot model in two dimensions is applicable to this problem. Fig.~\ref{allphases2d} below is the phase diagram of the $d=2$ Mandelbrot model, assembled from the observations of the preceding subsection.

\begin{figure}[h!]
\begin{center}
\includegraphics[scale=0.48]{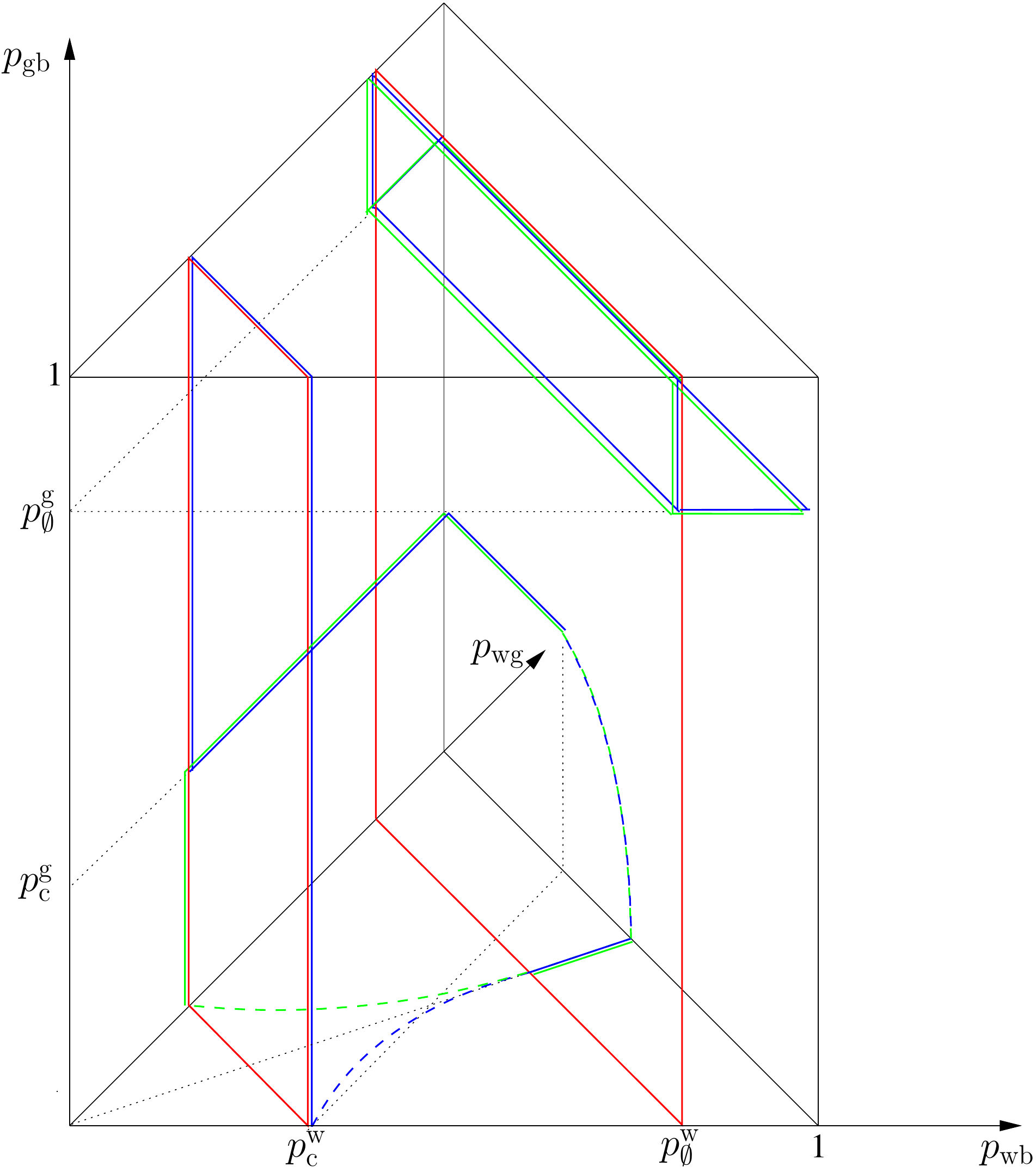}
\caption{The complete phase diagram of the $d=2$ three-vacuum system. Boundaries of the topological phases of white regions are marked in red; moving away from the origin, they contain WXC, white islands and no white at all (the aborted phase). Boundaries of the topological phases of grey regions are marked in green; moving down from the top, they contain no grey at all (the aborted phase), grey islands, and GXC. Boundaries of the topological phases of black regions are marked in blue; moving away from the origin, they contain black islands, BXC, and all space (the aborted phase). I regret the inadvertent chromatic confusion.}
\Label{allphases2d}
\end{center}
\end{figure}

Recall that in $d=2$ vacua of a given color may be found in three phases, wherein regions of that color form crossing curves, isolated islands, or are absent altogether. Consequently, the phase diagram contains two boundaries of each color (vacuum). The domain of the diagram is the product of a 2-simplex ($0 \leq  p_{\rm wb} + p_{\rm wg} \leq 1$) and a 1-simplex ($0 \leq p_{\rm gb} \leq 1$).

The following seven phases are distinguished in the diagram in Fig.~\ref{allphases2d}:

\begin{tabular}{|l|l|l|c|l|l|l|c|l|l|l|}
\hline
White & Grey & Black & & White & Grey & Black & & White & Grey & Black \\
\hline
WXC & islands & islands & & islands & GXC & islands & & $\emptyset$ & GXC & islands \\
\hline
& & & & islands & islands & islands & & $\emptyset$ & islands & BXC \\
\hline
& & & & islands & islands & BXC & & $\emptyset$ & $\emptyset$ & whole \\
\hline
\end{tabular}

\noindent
Interesting is the emergence of a grainy phase, in which observers inside the fiducial bubble see on their sky three types of vacua, none of which percolates. Also note that the boundary between the white-aborted phases containing GXC and BXC is marked in Fig.~\ref{allphases2d} with a continuous line. Its exact location is fixed by symmetry.

\subsection{Three vacua in three dimensions}
\label{3ddiag}

\begin{figure}[h]
\begin{center}
\includegraphics[scale=0.48]{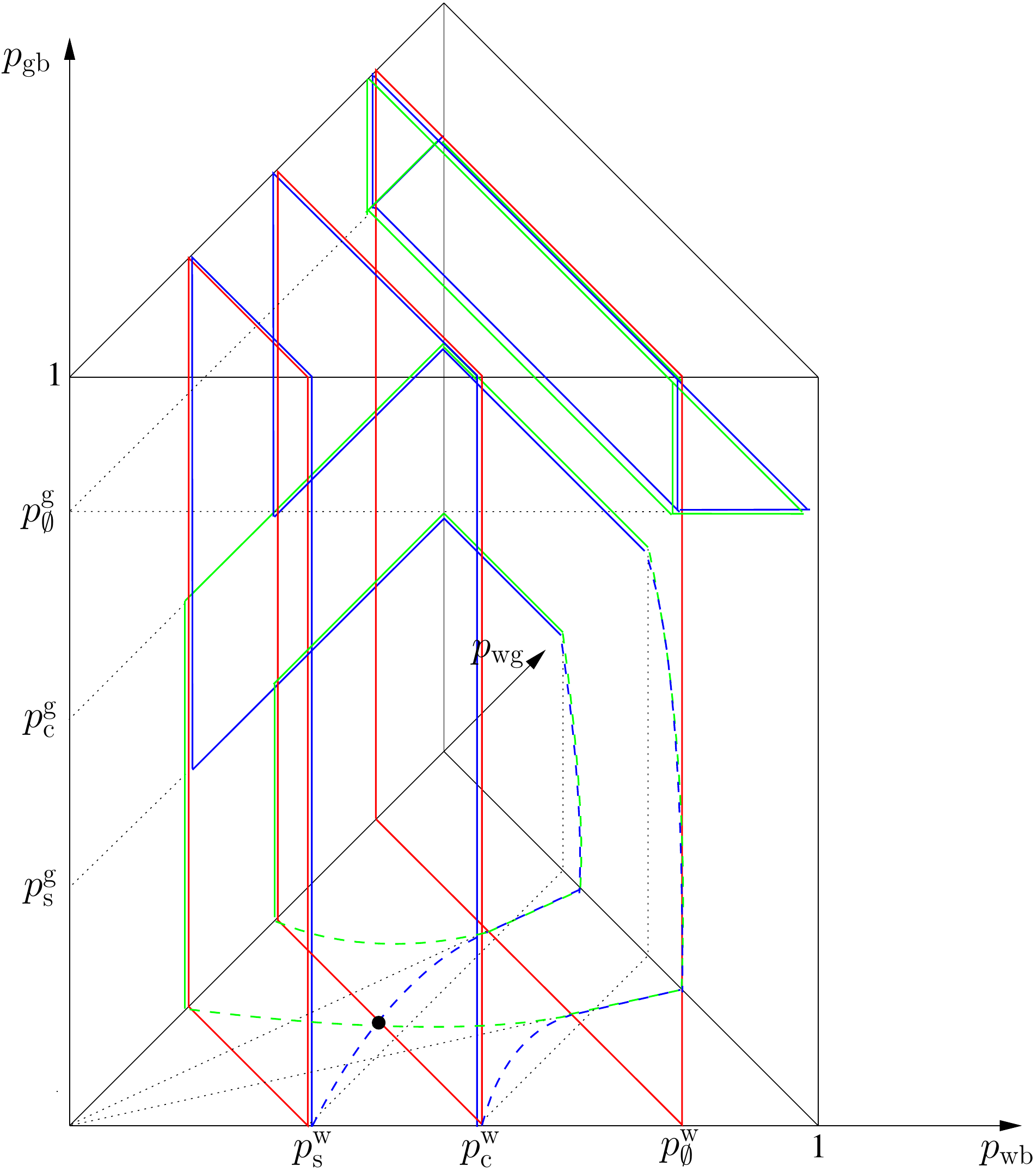}
\caption{The complete phase diagram of the $d=3$ three-vacuum system. Again, boundaries of white, grey, and black regions are marked in red, green, and blue, respectively. Moving away from the origin, the white regions in the successive phases contain WXS, WXC, white islands and no white at all (the aborted phase). Moving down from the top, the grey phases contain no grey at all (the aborted phase,) grey islands, GXC, and GXS. Moving away from the origin, the black regions contain black islands, BXC, BXS, and all space (the aborted phase). The black dot is explained in the text.}
\Label{allphases}
\end{center}
\end{figure}

The Mandelbrot model relevant for studying the topology of the inflating universe is in $d=3$. Its phase diagram, assembled from facts listed in Sec.~\ref{ingredients}, is presented in Fig.~\ref{allphases}. In three dimensions vacua of a given color may be found in four phases, in which regions of that color form crossing surfaces, crossing curves, isolated islands, or are absent altogether. Thus, the phase diagram contains three boundaries of each color (vacuum). 

The point marked with a black dot in Fig.~\ref{allphases} is established as follows. A priori, there are three possibilities for where the boundaries of the phases containing BXC and GXC intersect on the plane $p_{\rm gb}=0$: below the line $p_{\rm wb}+p_{\rm wg} = p^{\rm w}_{\rm c}$, on that line or above it. In the first case, the phase containing crossing curves of all three colors would persist to $p_{\rm gb}=0$. This is unlikely, because Point~3 of Grey Regions in Sec.~\ref{ingredients} identified non-zero $p_{\rm gb}$ as responsible for giving rise to that phase. I expect that the phase containing all three types of crossing curves persists only to some finite value of $p_{\rm gb}$, but a formal proof of this statement is outside the scope of this note. Next, if the boundaries of the phases with BXC and GXC met on $p_{\rm gb}=0$ above the line $p_{\rm wb}+p_{\rm wg} = p^{\rm w}_{\rm c}$, then another phase would exist, in which there would be no crossing curves of any color. But the absence of white crossing curves would imply that the combined grey and black regions form crossing surfaces, and the $d=2$ two-vacuum Mandelbrot model reveals that any such crossing surface must contain either a grey or a black crossing curve. Hence this possibility is also excluded. This implies that the phases with GXC and BXC meet on the plane $p_{\rm gb}=0$ precisely on the line $p_{\rm wb}+p_{\rm wg} = p^{\rm w}_{\rm c}$.

In summary, Fig.~\ref{allphases} contains the following fourteen phases:

\begin{tabular}{|l|l|l|c|l|l|l|c|l|l|l|}
\hline
White & Grey & Black & & White & Grey & Black & & White & Grey & Black \\
\hline
WXS & islands & islands & & islands & GXS & islands & & $\emptyset$ & GXS & islands \\
\hline
WXC & GXC & BXC & & islands & GXC & islands & & $\emptyset$ & GXC & BXC \\
\hline
WXC & GXC & islands & & islands & GXC & BXC & & $\emptyset$ & islands & BXS \\
\hline
WXC & islands & BXC & & islands & islands & BXC & & $\emptyset$ & $\emptyset$ & whole \\
\hline
WXC & islands & islands & & islands & islands & BXS & & & & \\
\hline
\end{tabular}

\noindent
The tubular phases arrange themselves in qualitatively novel ways. Recall that in the two-vacuum system crossing curves of both colors necessarily coexist. In contrast, the three-vacuum model contains three new, monochromatic tubular phases as well as a democratic tubular phase, which contains crossing curves of all three colors. The latter phase has an analogue for any number of vacua: the sequence of transition probabilities $p^{\rm i}_{\rm s} \leq p_{i\,i+1} < p^{\rm i}_{\rm c}$ (with all other $p_{\rm ij}$ set to zero) leads to a series of tubular networks, which are contained in one another in the order in which they nucleate.

\section{Discussion}
\Label{discussion}

This note determines the phase structure of the three-vacuum Mandelbrot model, which is a natural discrete toy model for studying the topology of three-vacuum eternal inflation. The model differs from those studied in \cite{ggv, freivogel1, freivogel2} in that it was not designed in order to reveal quantitative results for a specific scenario. Instead, it illustrates the wealth of behaviors, which can arise in a complex landscape containing many disorderly arranged minima: already with three vacua, the model has fourteen topological phases. Some of them are qualitatively novel, especially the all-tubular phase containing crossing curves of all colors, one inside another. These new phases are not artifacts of the discreteness of the model or of limiting the landscape to three vacua. On the contrary, their emergence is a good starting point for imagining the phase structure of realistic eternal inflation.

It is useful to focus first on the three monochromatic tubular phases, which contain crossing curves of only one color. It is natural to conjecture that with four vacua, the three-dimensional Mandelbrot model develops a grainy phase devoid of any crossing curves. (Indeed, such a phase exists in the three-vacuum Mandelbrot model in two dimensions.) This phase should have an analogue in realistic eternal inflation, at least when the number of vacua is large enough. Furthermore, it is clear that as the number of vacua grows, the grainy phase takes up more and more volume of the phase diagram. This is because additional vacua take up space, so it is increasingly difficult to sustain extended structures of any color. At the same time, the conditions for the existence of crossing curves and surfaces become increasingly limiting: if a vacuum $i$ decays to $n$ descendants, crossing curves or surfaces require the smalless of a sum of $n$ transition rates. Thus, one is led to the conclusion that as the number of vacua grows, the system is increasingly likely to find itself in the grainy phase, in which isolated bubbles of different vacua fill the geometry without forming crossing curves or surfaces.\footnote{This finding should not be compared with the results of \cite{ggv, freivogel1}, who were interested in the number of collisions in the past lightcone of a particular observer.} Of course, this statement relies on the landscape being so large as to be essentially random; the grainy phase is generic in the ensemble of all large landscapes.

The grainy phase is physically interesting. It is characterized by continuous cracking: a process in which the boundary of the parent vacuum region becomes disconnected. Ref.~\cite{susskind} concluded that in the two-vacuum model cracking leads to a formation of singularities in the surrounding descendant vacuum. In the many-vacuum case, however, every connected component of the parent vacuum region is surrounded by a collection of bubbles, each filled with a different vacuum. It is difficult to imagine surrounding such a structure with a homogeneous horizon. It would be exciting to study this scenario in detail.

\begin{figure}[t]
\begin{center}
\includegraphics[scale=0.29]{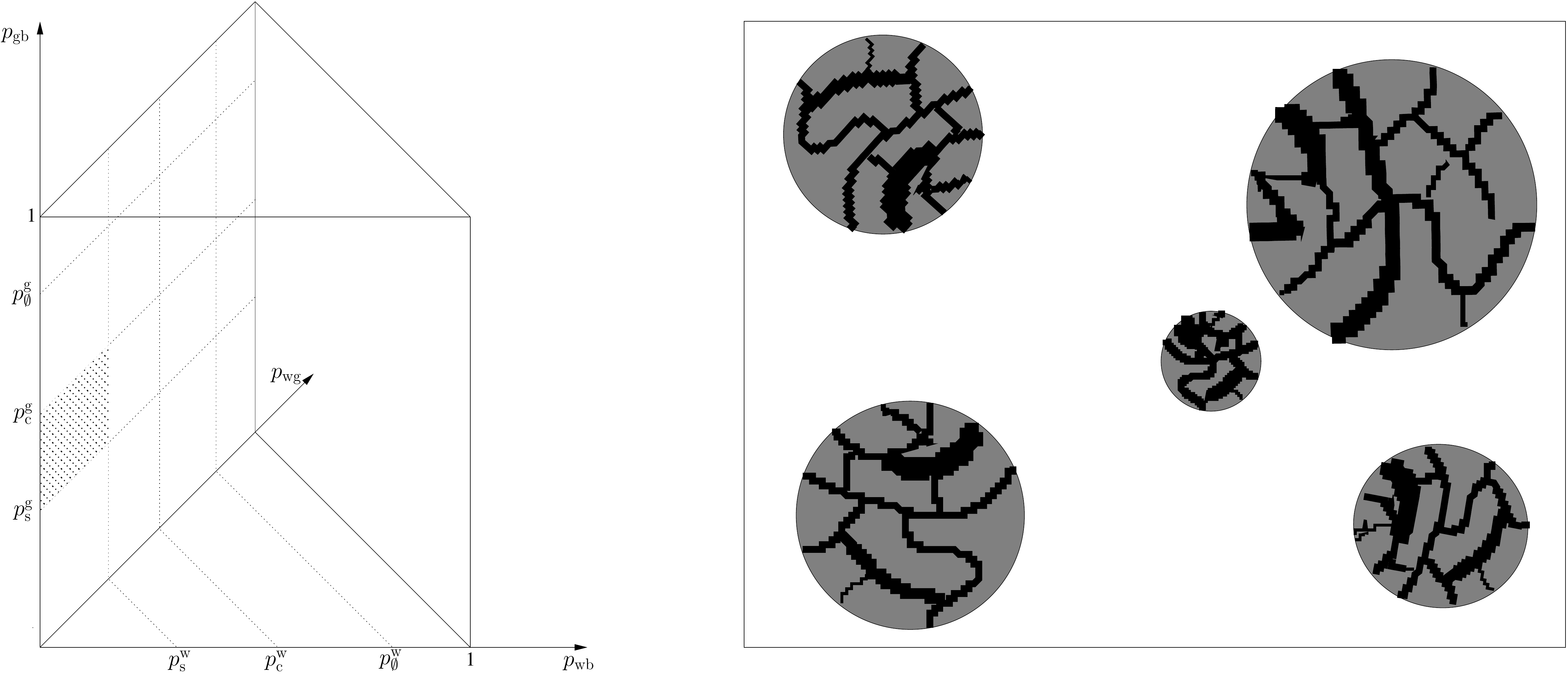}
\caption{The parameter range in phase space (left) that gives rise to global geometries containing grey islands, each of which is percolated by intrabubble BXC (right).}
\Label{hierarchy}
\end{center}
\end{figure} 

If the grainy phase is generic, can we forget all about the fourteen phases of the three-vacuum Mandelbrot model? Definitely not. Notice that Fig.~\ref{allphases} describes the existence of crossing curves and surfaces that span the infinite lattice that is the domain of the model. It does not tell us about curves, which cross the interior of only a single bubble of a given color. One example of this is shown in Fig.~\ref{hierarchy}: the range of parameters in the shaded region of the phase diagram gives rise to an arrangement containing grey islands, each of which is percolated by a black tubular network that could be called an intrabubble BXC.\footnote{This example was also considered in Sec.~VIII of \cite{susskind}.} The right panel of Fig.~\ref{hierarchy} presents a cartoon of the resulting geometry. 

The emerging picture is the following. On a global slice the grainy phase is generic. The global geometry becomes partitioned into randomly colored solids, which do not form crossing curves. However, particular spheres can contain interesting topologies such as intrabubble crossing curves described in the last paragraph, or intrabubble crossing surfaces. Moreover, from Sec.~\ref{3ddiag} we know that crossing curves can in principle be nested inside one another ad infinitum. Overall, the topological phases studied in the present paper and in \cite{susskind} are also generic in eternal inflation. Their appearance, however, is likely limited to a scale of several generations, beyond which the multiverse finds itself in the grainy phase. It would be extremely interesting to derive this picture of eternal inflation from a theoretical framework such as \cite{holographic}. 

\section*{Acknowledgements}
I thank Vijay Balasubramanian and Klaus Larjo for invaluable discussions of the material and the manuscript. I also thank Justin Khoury, Thomas S. Levi, Kris Sigurdson, Eva Silverstein and Mark Trodden for discussions. 
I thank the University of Pennsylvania for hospitality and opportunity to present this work. 
This project was supported by Natural Sciences and Engineering Research Council of Canada.

\end{document}